\documentclass[useAMS,usenatbib,usegraphicx,twocolumn]{mn2e}
  
\usepackage{float}
\usepackage{aas_macros}
\usepackage{threeparttable}
\usepackage{grffile}
\usepackage{amsmath, amssymb}
\usepackage{multirow}
\usepackage{wallpaper}
\usepackage{mathbbol}
\usepackage{rotating}
\usepackage{blindtext}
\usepackage[T1]{fontenc}
\usepackage{changes}

\usepackage{tocloft}
\newcommand{\ita}{\textit}

\newcommand{\beq}{\begin{equation}}
\newcommand{\eeq}{\end{equation}}  
\newcommand{\RNum}[1]{\uppercase\expandafter{\romannumeral #1\relax}}

\newcommand{\cm}{\mathrm{cm}}
\newcommand{\kpc}{\,\mathrm{kpc}}
  \title[Magnetised Galaxies]{The Origin of Filamentary Star Forming Clouds in Magnetised Galaxies}
  
  \author[B.~K\"ortgen]{Bastian~K\"ortgen$^1$, Robi~Banerjee$^1$, Ralph~E.~Pudritz$^{2,3}$ and Wolfram~Schmidt$^1$\\
  $^{1}$ Hamburger Sternwarte, Universit\"at Hamburg, Gojenbergsweg 112, D-21029 Hamburg, Germany \\
  $^{2}$ Department of Physics and Astronomy, McMaster University, Hamilton, ON L8S 4K1, Canada \\
  $^{3}$ Origins Institute, McMaster University, Hamilton, ON L8S 4K1, Canada  
  }

\date{Released 2018}

\pagerange{\pageref{firstpage}--\pageref{lastpage}} \pubyear{2016}

\begin{document}

\label{firstpage}
\maketitle

\begin{abstract}
Observations show that galaxies and their interstellar media are pervaded by strong magnetic fields with energies in the diffuse component being at least comparable to the thermal and even as large or larger than the turbulent energy. Such strong magnetic fields prevent the formation of stars because patches of the interstellar medium are magnetically subcritical. Here we present the results from global numerical simulations of strongly magnetised and self-gravitating galactic discs, which show that the buoyancy of the magnetic field due to the Parker instability leads at first to the formation of giant filamentary regions. These filamentary structures become gravitationally unstable and fragment into $\sim10^5 M_{\odot}$
clouds that attract kpc long, coherent filamentary flows that build them into GMCs. Our results thus provide a solution to the long-standing problem of how the transition from sub- to supercritical regions in the interstellar 
medium proceeds.
\end{abstract}

\section{Introduction}
One of the most fundamental questions in astrophysics is how do stars form.  Observations clearly show that this process takes place  in self-gravitating, turbulent, magnetized filamentary  molecular clouds \citep{Blitz07,Dobbs14,Andre13}.   
However, the question of how molecular clouds themselves arise is still largely unanswered. \\
\indent A major obstacle to assembling dense star forming gas from the diffuse interstellar medium is that magnetic fields dominate the energy budget  within the latter \citep{Beck01, Heiles05, Beck12}.    Star forming molecular clouds, on the other hand, are known to have gravitational energies that are typically a factor of 2-3 times greater than the magnetic energies \citep{Crutcher10,Crutcher12}.   This is often described by saying that diffuse H\textsc{i} clouds are magnetically subcritical whereas the molecular cloud cores are supercritical (as measured by the ratio of the gravitational to magnetic energies = the mass to flux ratio, $\mu$).  Because magnetic energy scales with gravity, by assuming a \ita{finite} mass reservoir and a nearly perfect conductivity of the interstellar medium,  a magnetically subcritical medium will remain so no matter how it is compressed, and this will prevent gravitational collapse.  \\
\indent This problem was recognized over sixty years ago by \cite{Mestel56}.  Their solution was that ambipolar diffusion (AD) in partially ionized gas might be able to mediate the transition from sub- to supercritical conditions.  It was soon pointed out that AD is not efficient enough to trigger the transition~\citep{Osterbrock61,Vazquez11a}.  The  problem of how magnetically supercritical cloud cores form out of a magnetically subcritical, diffuse medium has remained a challenge ever since  \citep[see also more recent numerical studies on this issue, e.g.][]{Vazquez11a, Koertgen15}.  \\
 \indent It is well known that an attractive potential solution is that magnetic fields are buoyant and will bubble out of sufficiently magnetized galactic disks, 
 resulting in gas flowing back to the biplane and concentration into clouds, i.e. 
 the Parker instability.   
 The density increase in the magnetic valleys is, for cases without self-gravity, only a factor of $\sim2$, too small to provide the 
seeds for potential further growth of molecular clouds \citep{Kim02}.  Parker instability and subsequent cloud formation could be triggered 
 by the passage of a compressive spiral wave however \citep{Blitz80,Elmegreen82} . 
The inclusion of disk gravity lead to the formation of $\mathcal{O}(10^5\,\mathrm{M}_\odot)$ H\textsc{i} clouds by the Parker instability \citep{Kim02,Mouschovias09}. However, more attention has focused instead on magneto-jeans-fragmentation \citep{Kim02}.  Although other processes -- such as cloud-cloud collisions and mergers \citep{Tasker09} and colliding warm neutral medium flows \citep{Ballesteros99b,Vazquez11a} have been simulated, these do not confront the question of magnetized galaxies.
Once a magnetically supercritical medium is established, the mass needed to assemble massive clouds could be collected by  converging motions from a large mass reservoir. 
The necessary column density of such H\textsc{i} connected regions is about $2\times 10^{21} \cm^{-2}$ and would require a catchment region of at least one $\kpc$ \citep{Mestel85}.\\ 
\indent In this work, we explore a major aspect of global Parker instability in 3D galaxies that has been overlooked.   Our simulations show that large scale Parker instabilities ranging over many kpcs in a galactic disk can create kpc long filaments.   Gas draining into such long magnetic valleys will ultimately exceed the criterion for the gravitational fragmentation 
of filaments.   Subsequent flows along these filaments towards these initial fragments result in the rapid growth of massive GMCs.   This addresses two crucial issues about star formation in magnetised galaxies, namely, how can star forming gas form out of an initial  magnetically dominated diffuse interstellar medium, and how do  filamentary GMCs arise as a consequence of this process? \\
\indent  We adopt the simplest possible model of a galaxy that does not have a companion galaxy driving strong spiral arms. Under these conditions, galactic shear will ensure that the initial magnetic field is overwhelmingly toroidal in structure. We perform state of the art numerical simulations of  an isolated, magnetized  galaxy and present 
our results on the appearance and consequences of the Parker instability in its initially subcritical interstellar medium.
\section{Numerics and Initial Conditions}
For our study of galaxy evolution we use the \textsc{flash} code \citep[v4.2.2,][]{Dubey08}. \\
\indent The disc is initialised in the center of a cubic box with edge length of $L=40$~kpc. The disc radius is set to \mbox{$R_\mathrm{disc}=10$~kpc}. For both the hydrodynamic (HD)  and magnetohydrodynamic (MHD) simulations, the initial gas density of the galaxy is 
a function of radius $R$ and height above the midplane $z$. Following \citet{Tasker09} and taking into account the Alfv\'{e}n velocity, the initial density profile is given as 
\begin{equation}
\rho(R,z)=\frac{\kappa c_\mathrm{s}\sqrt{1+\frac{2}{\beta}}}{\pi G Q_\mathrm{eff}H(R)}\mathrm{sech}^2\left(\frac{z}{H(R)}\right).
\end{equation}
Here, $Q_\mathrm{eff}$ is the effective Toomre parameter, $\beta=2c_\mathrm{s}^2/v_\mathrm{a}^2$ the ratio of thermal to magnetic pressure, $\kappa$ the epicyclic frequency and $H(R)=R_\odot(0.00885+0.01719R/R_\odot+0.00564(R/R_\odot)^2)$ the scale height at distance $R$ from the center and $R_\odot=8.5\kpc$ the solar distance. This realisation essentially yields a thin disc with $H(R_\odot)\sim270\,\mathrm{pc}$. For numerical reasons, we define the innermost and outer part as being initially gravitationally stable and set $Q_\mathrm{eff}=20$ for radii $R<2$~kpc and $R>8.5$~kpc. Throughout the main disc the initial value for $Q_\mathrm{eff}$ is set at $Q_\mathrm{eff}=2$. In addition to self-gravity we use a fixed, stationary external logarithmic potential 
of the form
\begin{equation}
\Phi_\mathrm{ext}=\frac{1}{2}v_0^2\mathrm{ln}\left[\frac{1}{R_c^2}\left(R_c^2+R^2+\left(\frac{z}{q}\right)^2\right)\right],
\end{equation} 
which takes into account old stars and dark matter and yields a flat rotation curve 
\begin{equation}
v_\mathrm{rot}\left(R\right) = v_0\frac{R}{\sqrt{R_c^2+R^2}}.
\end{equation}
For our magnetised galactic disc model, we begin with an initial state in which the magnetic field is entirely toroidal. We scale the magnetic field strength with the gas density as $B\propto\sqrt{\rho}$ to achieve a constant ratio of thermal to magnetic energy density  ( $ \equiv \beta $ ) 
throughout the disc \citep{Beck15}. Our fiducial value is $\beta=0.25$, which gives $B_0(R=R_\odot)\sim30\,\mu\mathrm{G}$. The medium outside the disc is weakly magnetised and rather tenuous and thus does not exert any significant thermal 
or magnetic pressure on the disc.   Our MHD simulations are initially magnetically dominated, with a subcritical mass-to-magnetic flux ratio of $\mu/\mu_\mathrm{crit}=0.45$,
where the critical mass to flux ratio is $ \mu_{crit} =0.16/\sqrt{G}\sim620\,\mathrm{g}^{1/2}\,\mathrm{s}\,\mathrm{cm}^{-1/2}$. \\
\indent The simulations are non-isothermal and we follow the recipe by \citet{Koyama02} to heat and cool the gas in the disc. The form of the cooling curve allows for a thermally unstable regime in the density range \mbox{$1\leq n/\mathrm{cm}^{-3}\leq10$}.\\
\indent The equations of ideal MHD are solved by using a HLL5R Riemann solver \citep[e.g.][]{Waagan11} and the self-gravity is treated with a Barnes-Hut tree solver \citep[optimised for GPU][]{Lukat16}. We use outflow boundary conditions for the 
MHD and isolated ones for the self-gravity. The root grid is at a resolution of \mbox{$\Delta x_\mathrm{root}=625\,\mathrm{pc}$} and we allow for five additional refinement levels, thus giving a peak spatial resolution of \mbox{$\Delta x_\mathrm{min}=19.5\,\mathrm{pc}$}. The numerical grid is refined once the local Jeans length is resolved with less than 32 grid cells and derefined when it exceeds 64 cells. On the highest level of refinement, we incorporate an artificial pressure to ensure that the Jeans length is refined with at least four grid cells.
\vspace{-0.5cm}
\section{Results}
\begin{figure*}
\begin{tabular}{lcl}
\includegraphics[height=0.24\textwidth]{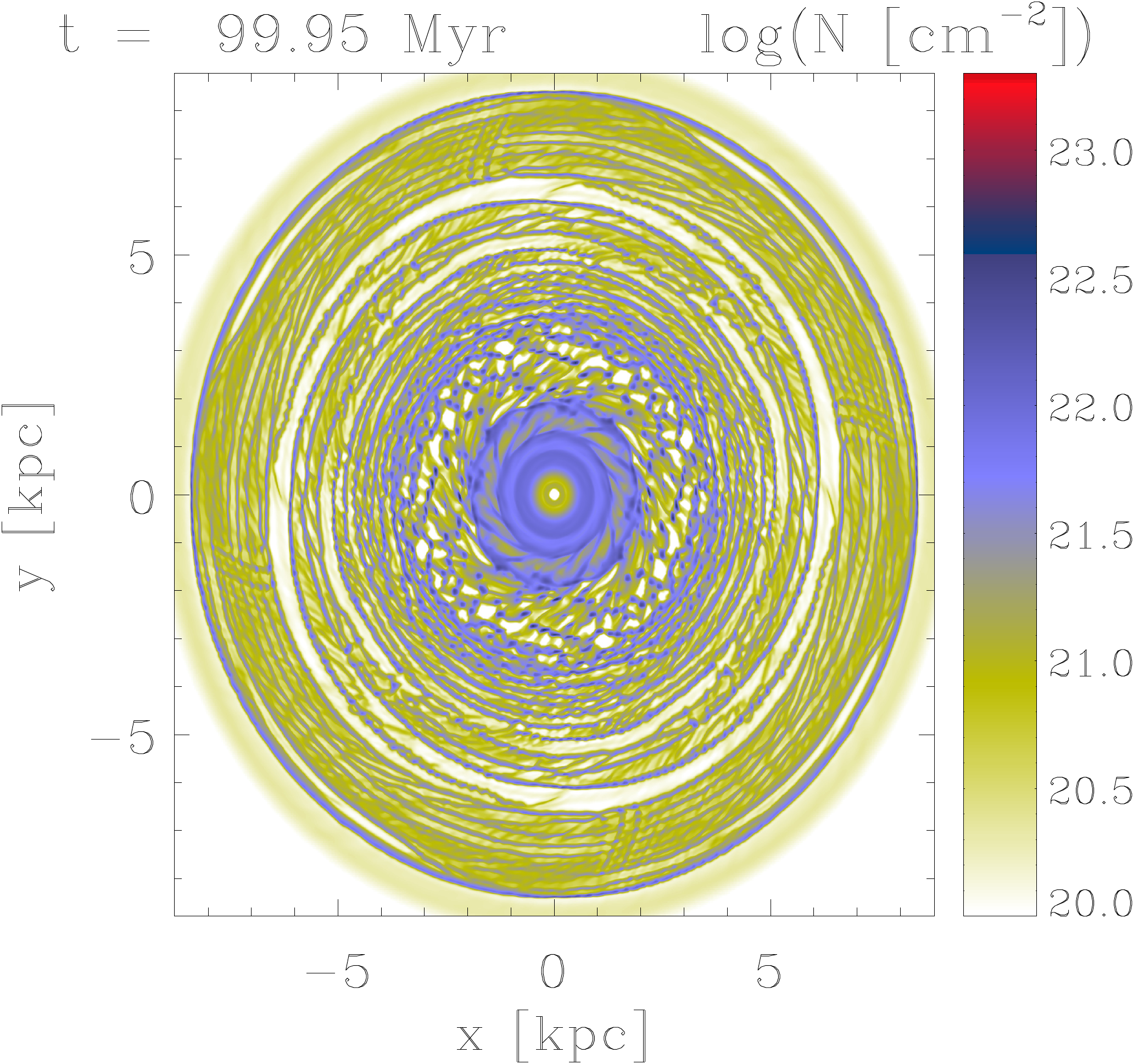}&\includegraphics[height=0.24\textwidth]{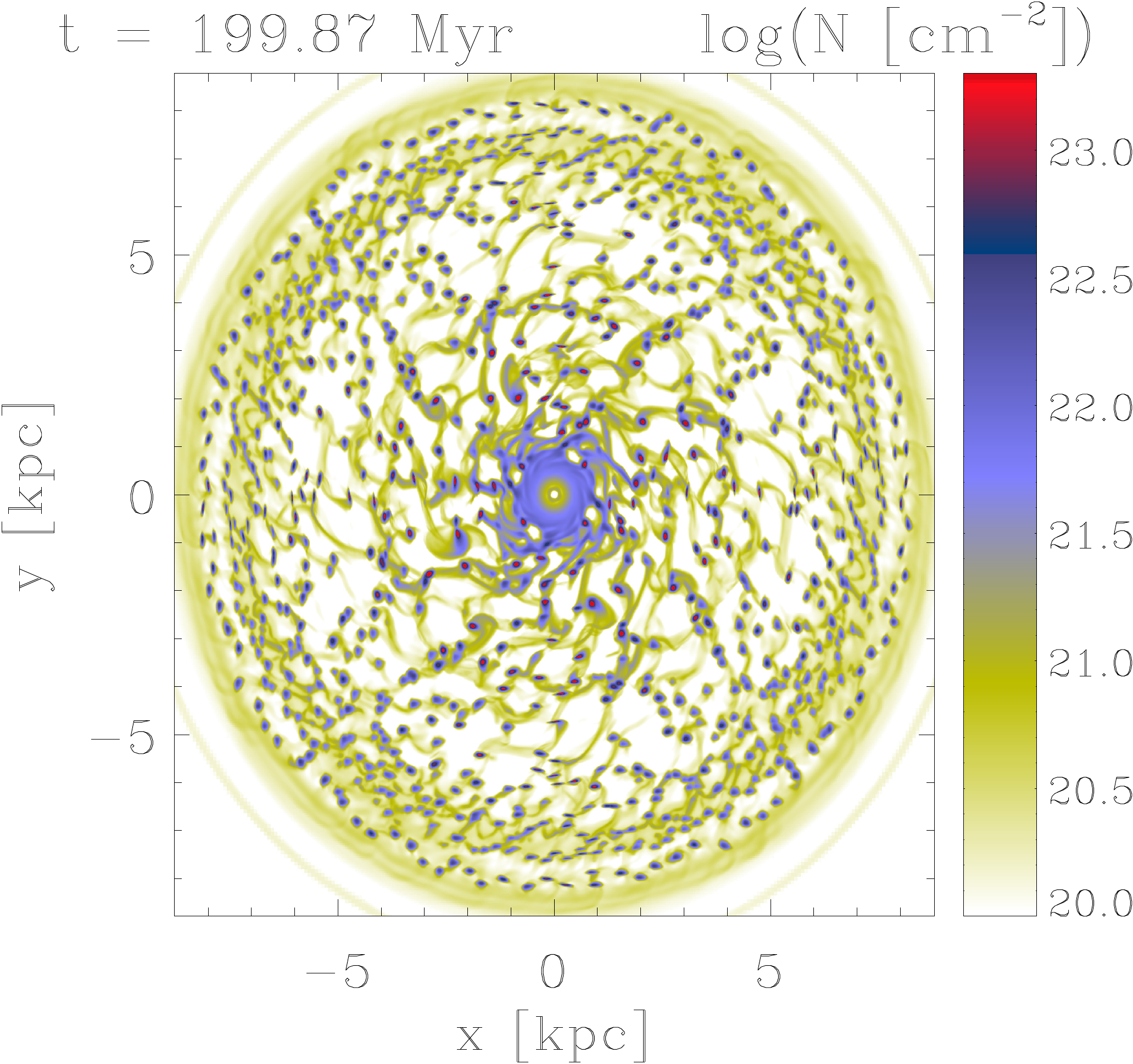}&\includegraphics[height=0.24\textwidth]{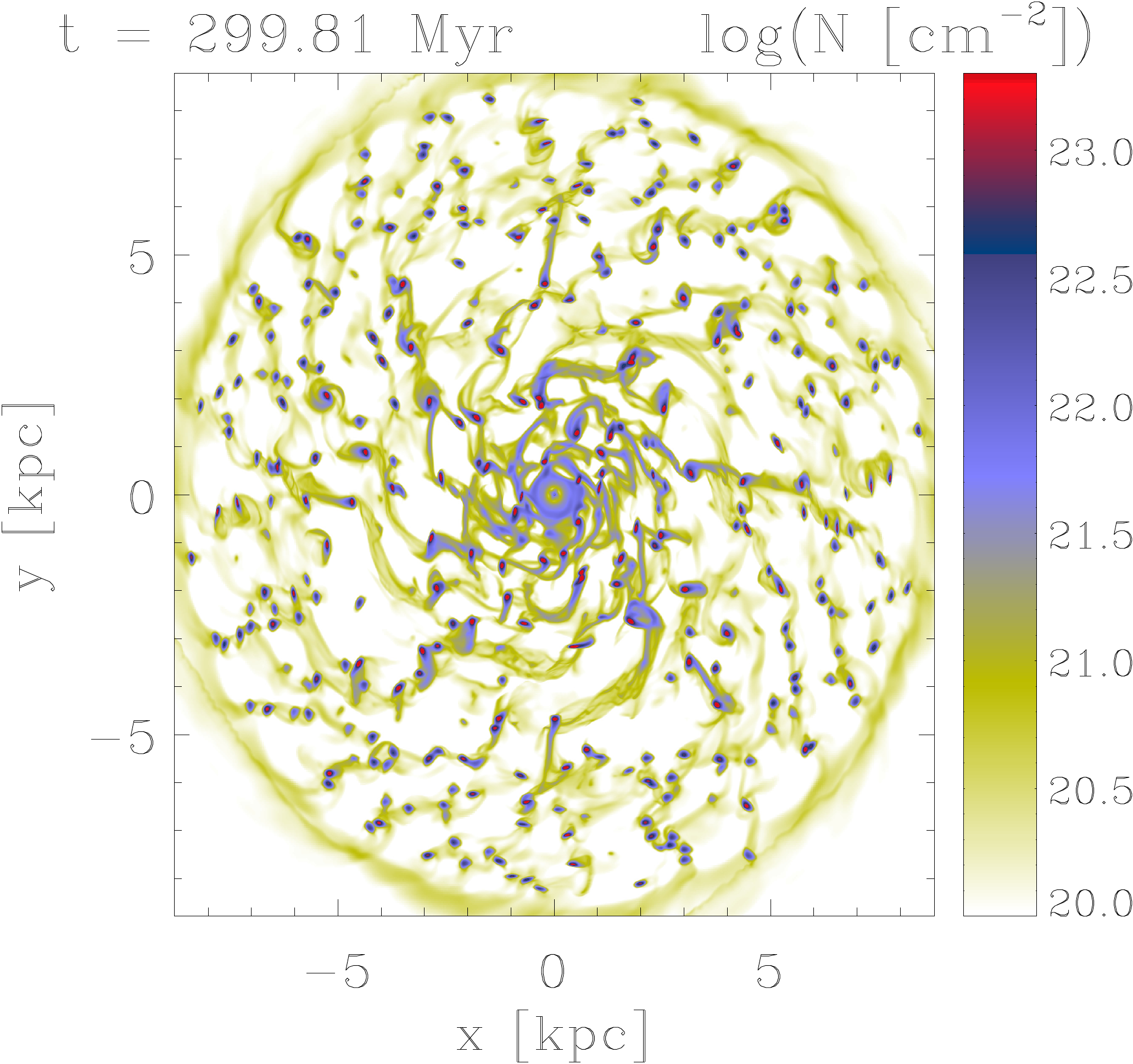}\\
\includegraphics[height=0.24\textwidth]{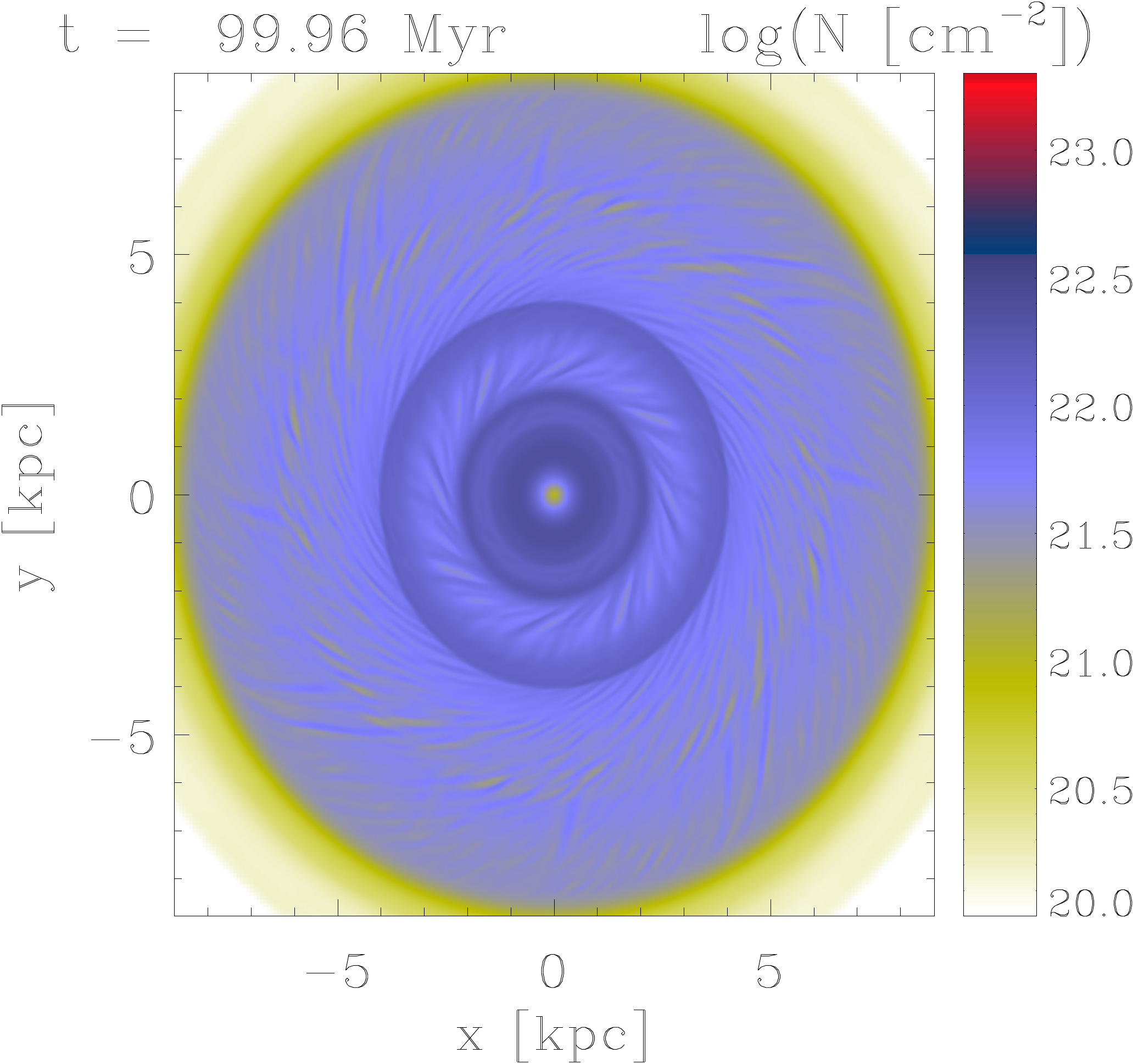}&\includegraphics[height=0.24\textwidth]{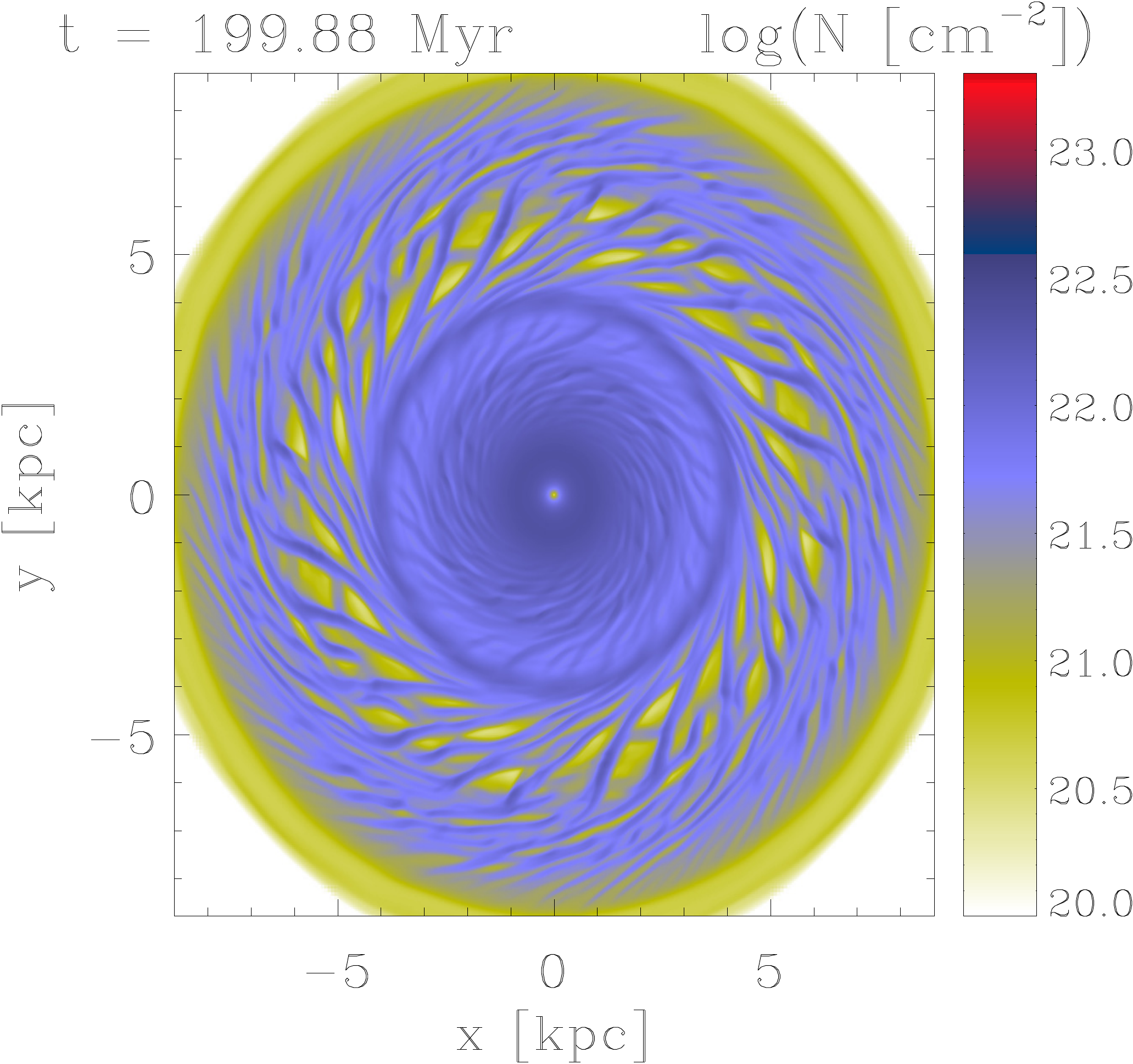}&\includegraphics[height=0.24\textwidth]{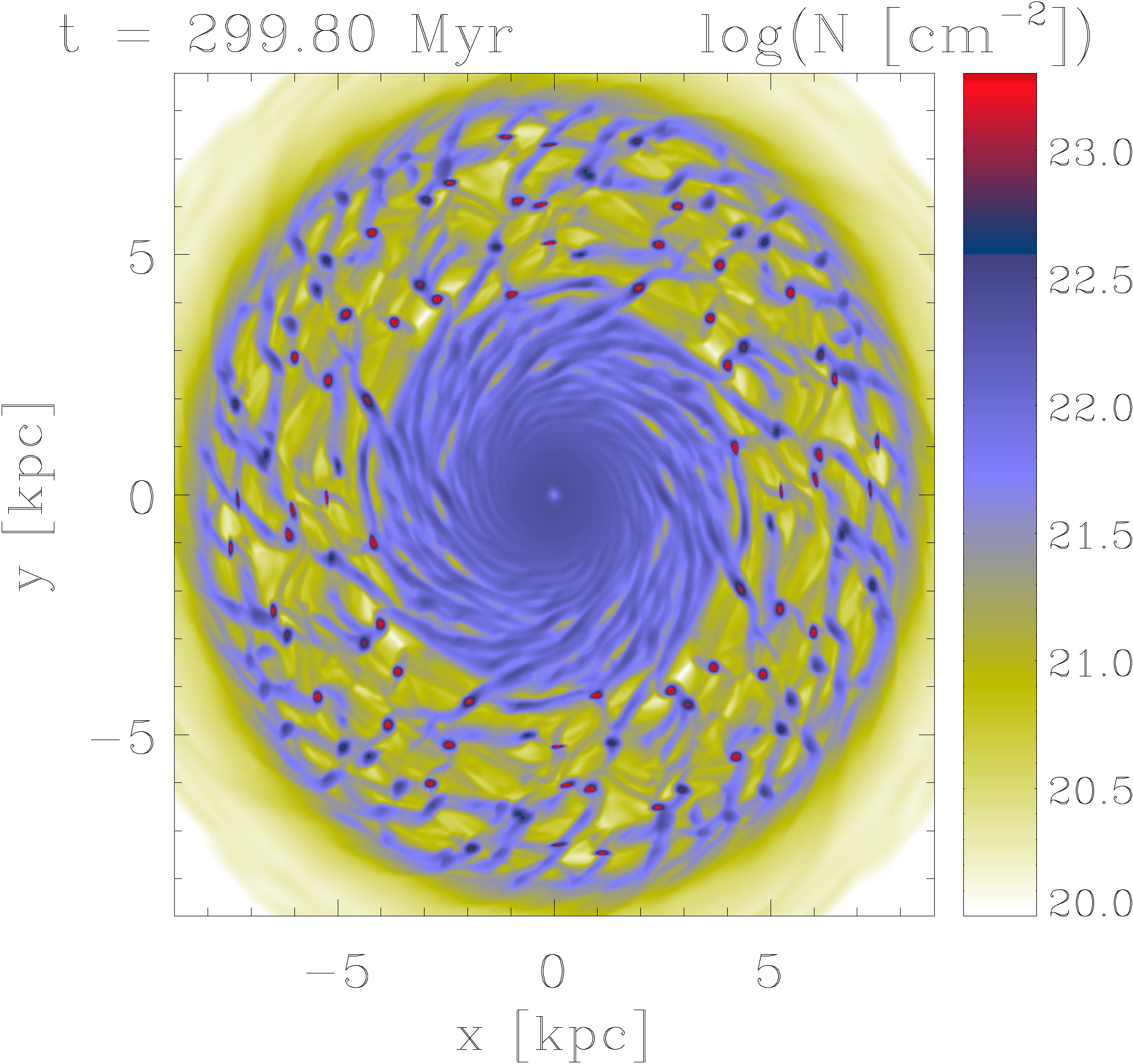}\\
\end{tabular}
\caption{ Column density maps  of the simulated galaxies. Top: HD control disc in time steps of 100 Myr. Bottom: Same for the case with MHD, characterised by a low plasma-$\beta$. The fragmentation pattern of the two discs (t=100 Myr for the HD case and t=200 Myr for the MHD disc) are significantly different. While in the HD case, the disc breaks up into ring-like structures as expected from classical perturbation theory, the MHD disc shows a pattern that is dominated by spurs which extent both in the azimuthal and radial direction. Movies of these two simulations can be found online.}
\label{fig1}
\end{figure*}

\begin{figure}
\centering
\includegraphics[height=0.30\textwidth]{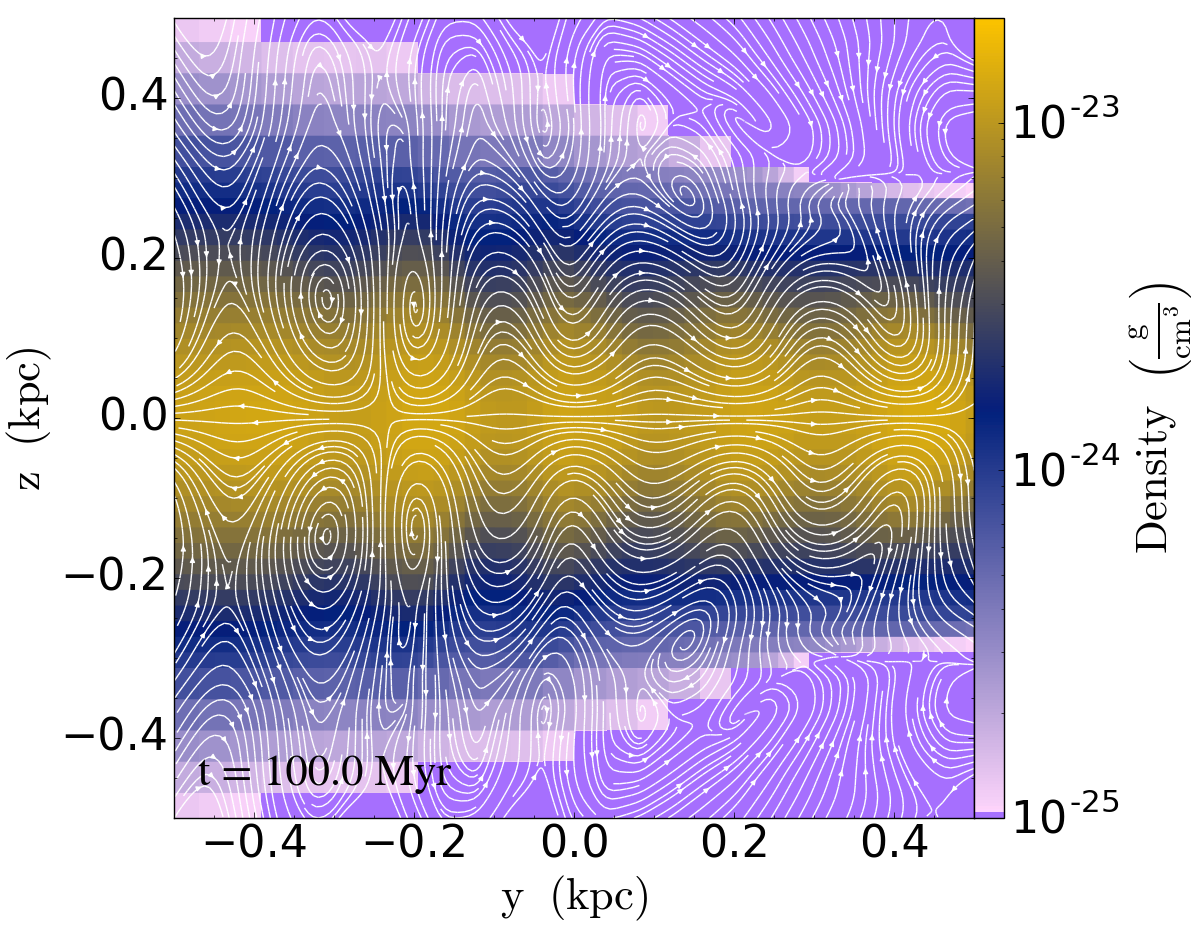}
\caption{Slice of density through the magnetised disc at \mbox{$t=100$~Myr}. The axes show the offset to the center of the region, which is at $y=+5\,$kpc and $x=z=0$. Overlaid in white are magnetic field lines. The Parker instability is clearly seen in its developed phase. The density enhancements correlate well with the emergence of magnetic valleys.}
\label{fig2}
\end{figure}

\begin{figure*}
\begin{tabular}{lcl}
\includegraphics[width=0.23\textwidth]{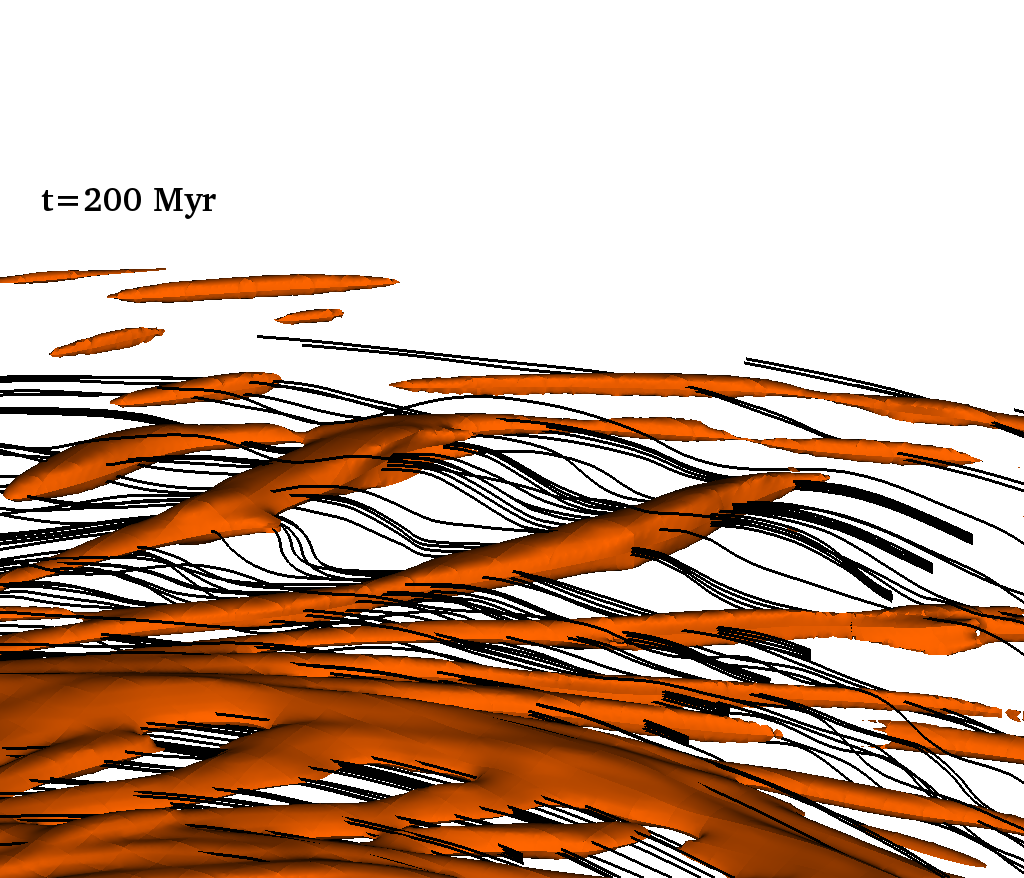}&\includegraphics[width=0.23\textwidth]{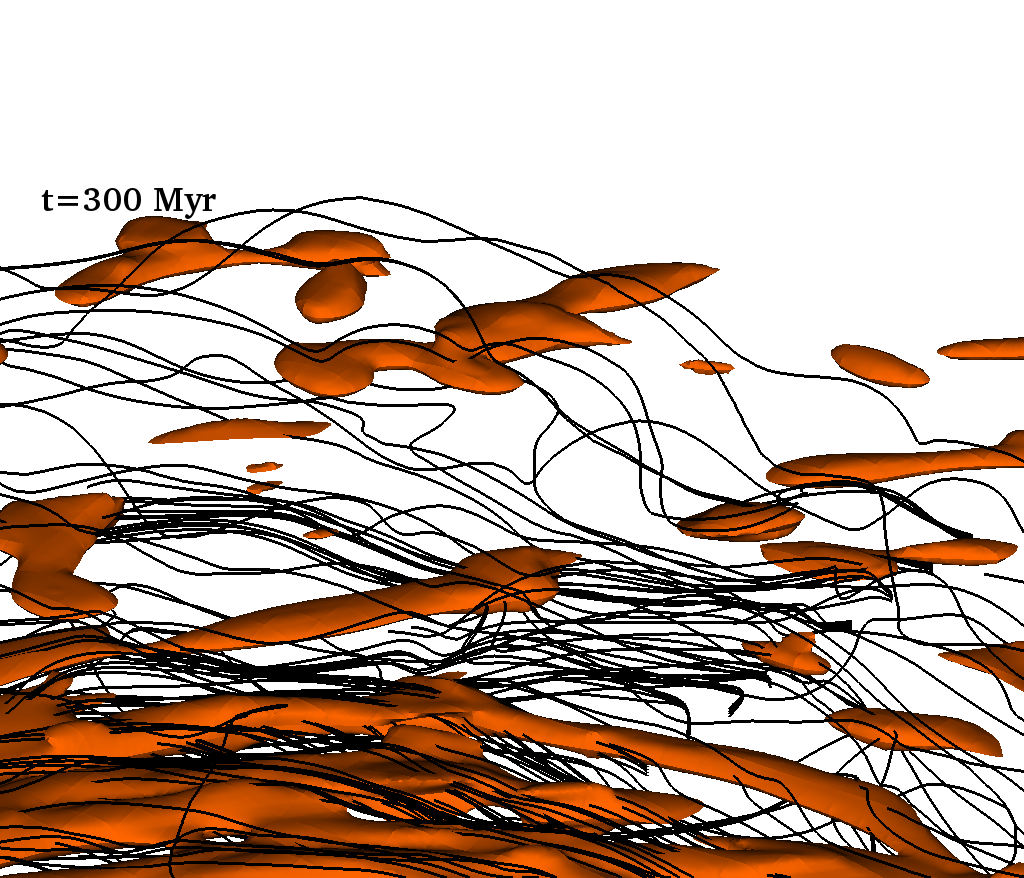}&\includegraphics[width=0.23\textwidth]{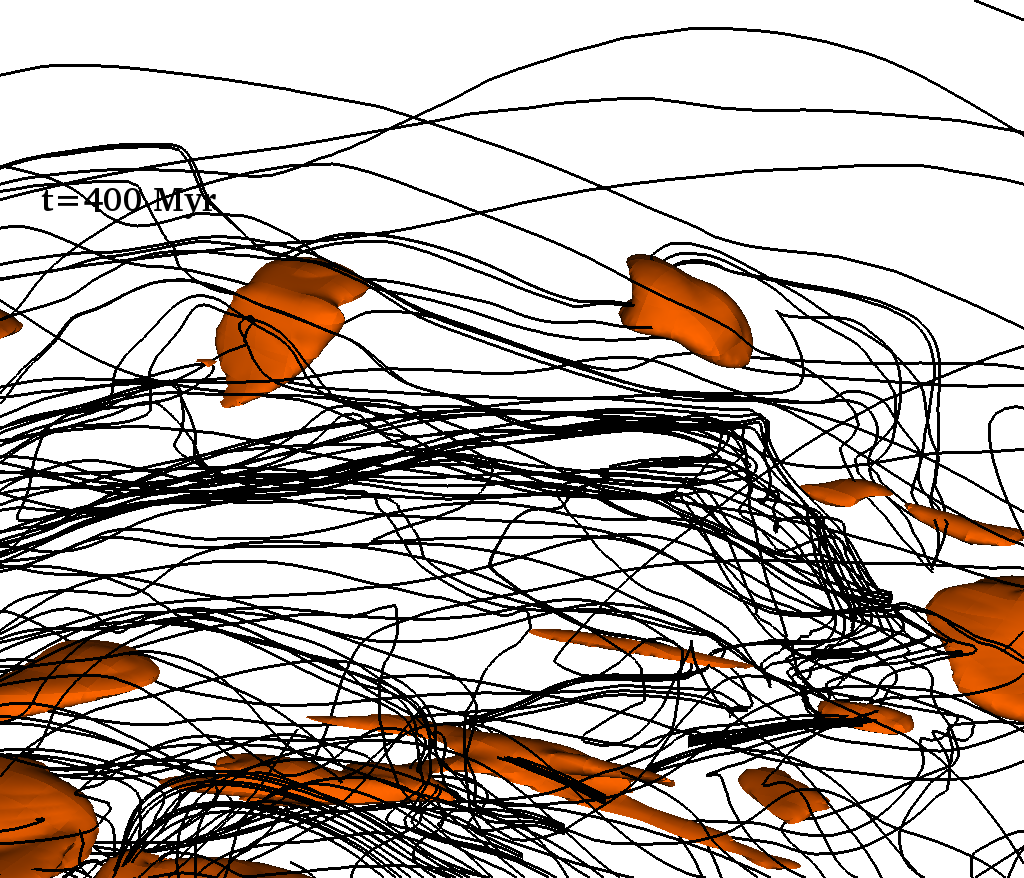}
\end{tabular}
\caption{Zoom-in to selected areas of the disc for the reference run ($\beta=0.25$), starting from the point of spur-formation in the outer parts of the disc. It is observed that the magnetic field lines buckle, typical for the Parker instability (left). Near the valleys, where the field lines buckle down, gas motions along the field lines form filamentary structures. At later times, when the disc has partially fragmented into individual clouds, the magnetic field lines are still locally (preferentially in the outer parts) buckled. From the middle panel it is evident that the formed clouds are initially magnetically connected. At late times the B-field becomes highly distorted.}
\label{fig3}
\end{figure*}

\begin{figure}
\begin{tabular}{ll}
\includegraphics[height=0.21\textwidth]{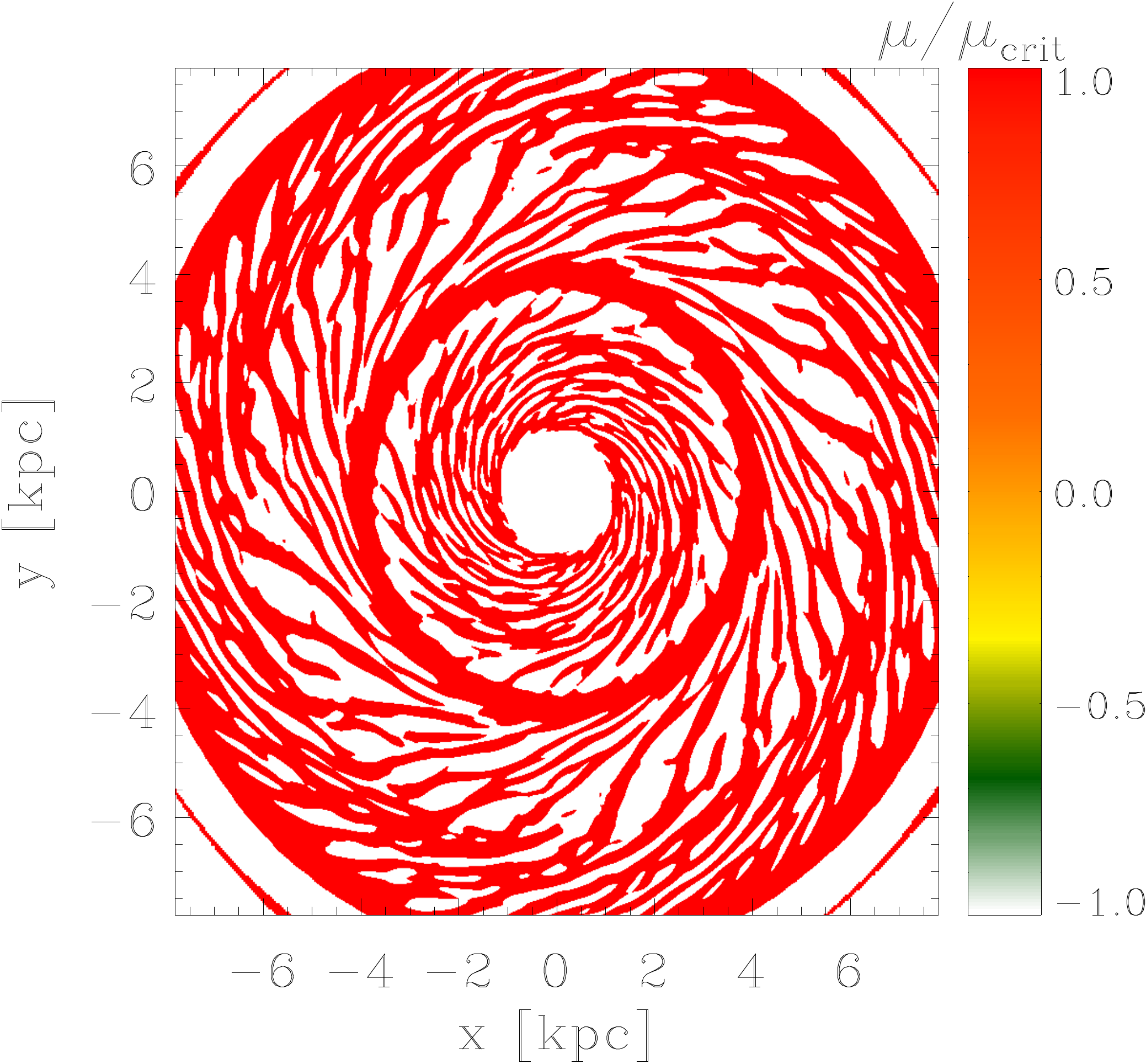}&\includegraphics[height=0.21\textwidth]{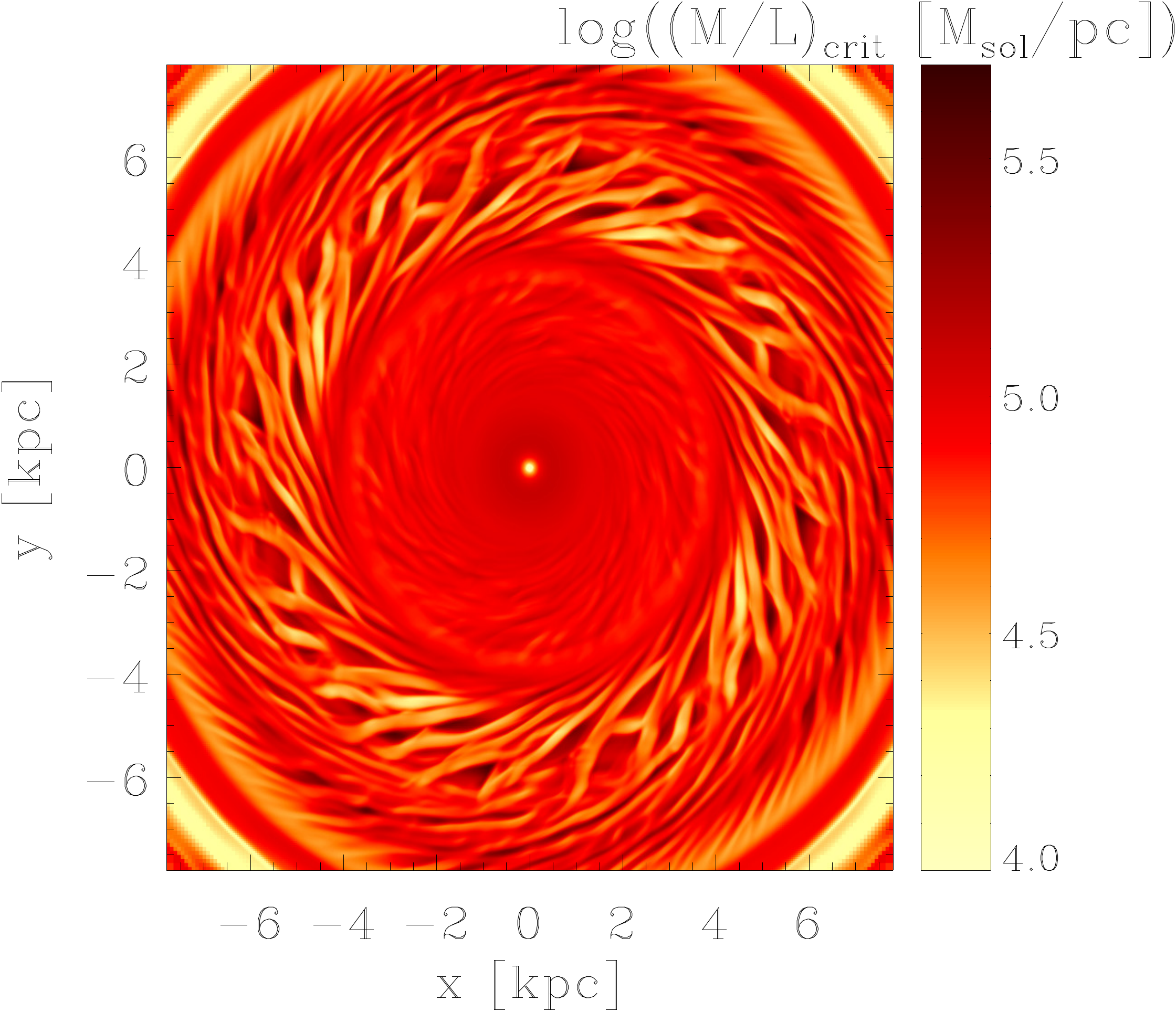}
\end{tabular}
\caption{The Parker Instability in the disc at $t=200\,\mathrm{Myr}$. \ita{Left:} Face-on view of the mass-to-flux ratio shown in 'binary mode' (supercritical equal to 1, red, and subcritical 
equal to -1, white). \ita{Right:} \ita{Critical} line-mass of the disc. The formed spurs are magnetically supercritical and more prone to gravitational fragmentation as indicated by their lower critical line-mass.}
\label{fig4}
\end{figure}

\begin{figure}
\centering
\includegraphics[width=0.285\textwidth,angle=-90]{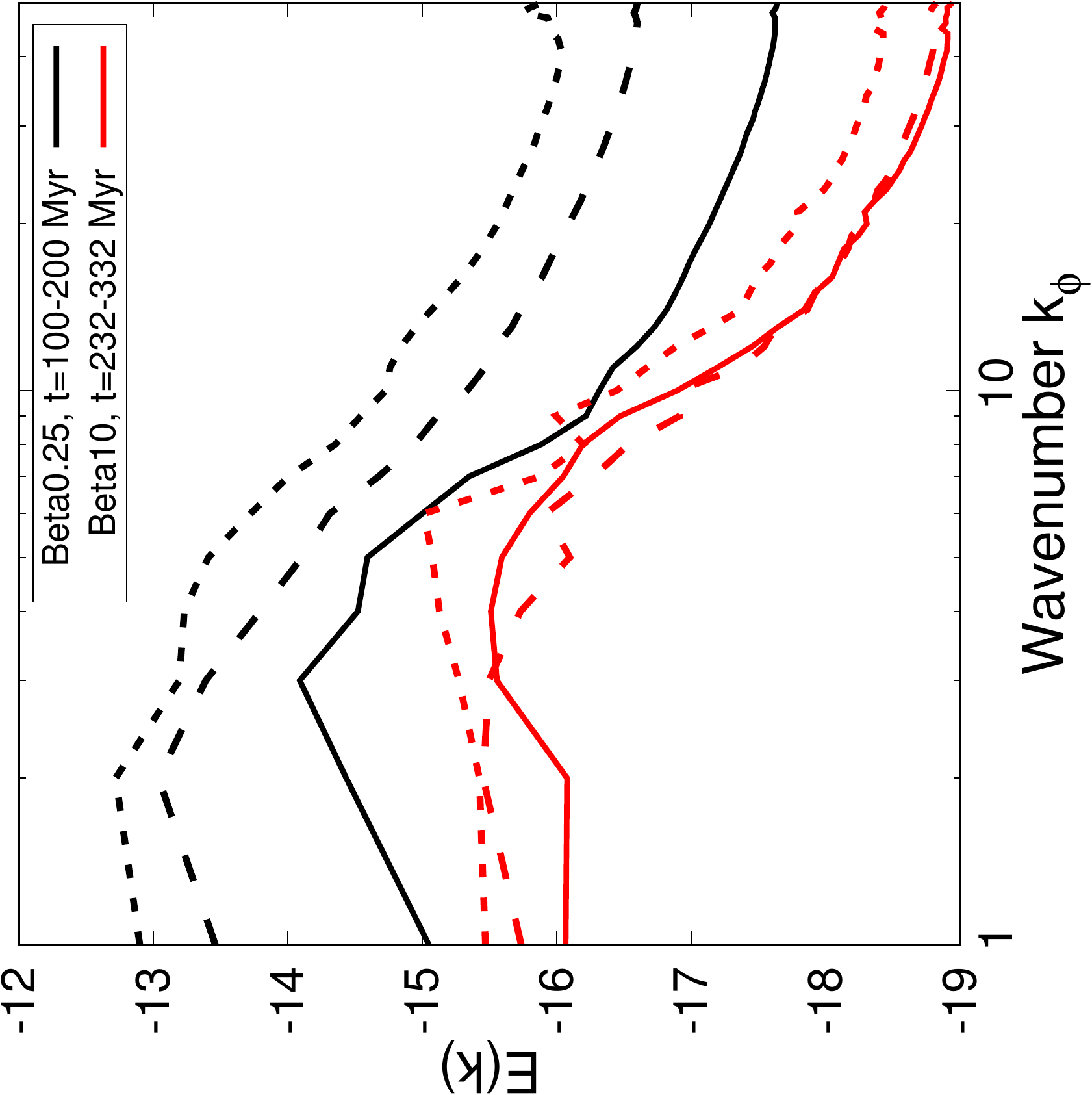}
\caption{Azimuthal energy spectra of the vertical magnetic field component, which is used as a measure of the distortion of the magnetic field. Different line styles depict different times (early: solid, late: short-dashed). The last time step is chosen such that the disc is still in the phase of linear instability and this region is hence representative for the entire disc.}
\label{fig5}
\end{figure}

The global evolution of the galaxies is shown in Fig. \ref{fig1} for two types of simulations - HD and MHD.  As a reference case, we first consider a HD disc galaxy, which is depicted in the upper row. The disc starts to fragment into rings at around $t=100$~Myr due to gravitational (Toomre) instability. 
The fragmentation is limited to the outer parts of the disc. 
At $t=200$~Myr the rings have fragmented into individual objects (henceforth clouds). 
At this stage, gravitational and tidal interactions between clouds start to become dominant, thereby inducing cloud-cloud collisions and mergers. The latter is clear from the snapshot at time $t=300$~Myr, where the number of clouds has been greatly reduced. \\
\indent The magnetised disc galaxy evolves quite differently.  The magnetic field is strong enough to 
suppress gravitational fragmentation of the disc which is readily seen at $t=100$~Myr, where the disc galaxy appears much smoother, while its hydrodynamic counterpart has already started to fragment. The surface density remains almost constant except for some small scale perturbations, which are indicative of a beginning instability. With time, the instability grows and filaments of increased density, which extend both in the azimuthal and radial direction, are formed. They are a few kiloparsec long and the medium between the filaments shows a decreased surface density. We also see that the spacing between those filaments becomes smaller when looking at the regions near the center of the galaxy. At $t=300$~Myr, the spurs in the outer part of the disc have fragmented into a large number of clouds. These clouds appear to be connected via filaments of high density, which are aligned in the same manner as the large scale spurs at earlier times. These results are still found even in simulations where we imposed an initial turbulent velocity of $\sim10\,\mathrm{km\,s}^{-1}$. \\
\indent In Fig. \ref{fig2} we show a density slice at $t=100\,\mathrm{Myr}$ in the YZ-plane, centered at $y=+5\,$kpc and $x=z=0$, with the magnetic field lines overplotted. The field lines are buoyant within 200~pc above/below the midplane and appear tangled at larger distances due to turbulence caused by accretion of matter onto the disc \citep{Klessen10}. The tangling of the magnetic field is prominent in this plane as this (radial) component, which is initially zero, is much weaker compared to the toroidal field. \\
\indent The density field is well correlated with the buoyancy of the field lines. Regions with vertically lifted field lines show a decreased density in contrast to the areas with magnetic valleys, where the density is increased. In the former regions, gas is pushed up and out of 
the disc by the rising field lines and then flows back down along these towards the magnetic valleys. The spacing of the valleys is observed to be around $\sim200\,\mathrm{pc}$ in the YZ-plane and $\sim400-600\,\mathrm{pc}$ in the XZ-plane (not shown), 
consistent with theoretical expectations, which give $\lambda_\mathrm{YZ}\sim H(R)$ and $\lambda_\mathrm{XZ}\sim(10-20)H(R)$ \citep[see also][]{Kim01}. \\
\indent Fig. \ref{fig3} shows both the iso-surface maps of the density ($n_\mathrm{min}=10\,\mathrm{cm}^{-3}$ in red) and magnetic field lines (black) at three distinct stages of the magnetised disc.  At $t=200$~Myr, the spurs are readily seen and magnetic field lines are perturbed, being dragged by buoyant flows in the vertical direction. 
This topology of the magnetic field lines is a typical for the Parker instability, where field lines buckle up and down in a vertically stratified disc. It is also seen that the Parker instability is most efficient in the outer parts of the disc, whereas fragmentation of the inner part is suppressed and the field is observed to be still rather unperturbed. \\
\indent We emphasize that the difference in these patterns  is due to the fact that the dynamical time of the Parker instability is shorter than that of the 
Toomre instability -  we deduce from our initial conditions that it is a factor of 2-3 shorter, with $t_\mathrm{PI}\sim90-100\,\mathrm{Myr}$ at $8\kpc$.  
At later times, $t=300$~Myr, the magnetic field lines become even more perturbed, while the filaments have fragmented into clouds with initial masses $\sim10^5\,\mathrm{M}_\odot$. 
Even at this stage, the field buoyancy can be clearly identified. 
Although the magnetic field becomes compressed in the magnetic valleys, the spurs proceed to fragment into clouds. 
The reason for this is that gas flows coherently along the field, converges and accumulates in the magnetic valleys, which increases the line-mass of the filaments \citep{Andre13}. As the line-mass increases, then so too will be the mass-to-magnetic flux ratio since the inflow does not drag field into the filament which is contrary to the scenario suggested by \citet[][see also Fig. \ref{fig4}]{Mestel85}.  Thus, the inflow into the filament pushes the system towards higher line-mass and renders it magnetically supercritical. At the 
same time, the filaments reveal low levels of turbulence and thus the lowest threshold line-mass for gravitational fragmentation. It is further seen that these clouds are connected by the same magnetic field lines. The field itself stays buoyant thereby further allowing less dense material to flow to the magnetic valleys. When the disc has fully fragmented at $t=400$~Myr (rightmost panel), the magnetic field becomes locally disordered as gravitational interactions between the clouds take hold.\\
We show  in Fig. \ref{fig4} the normalised mass-to-magnetic flux ratio $\mu/\mu_\mathrm{crit}$. It is depicted in 'binary mode', where supercritical regions are shown in red and subcritical areas in white, respectively. 
Note that the spurs appear as magnetically supercritical regions in the mass-to-flux ratio map. Regions, where the magnetic field lines are lifted out of the disc appear magnetically subcritical due to the greatly reduced gas column density. \\
\indent The gravitational stability of filaments is measured in terms of  
the critical gravitational  line-mass \mbox{$ (M/L)_\mathrm{crit}=2\left(\sigma^2_\mathrm{rms}+v_\mathrm{A}^2+c_\mathrm{S}^2\right)/G$} \citep{Fiege00a}, which is shown in the right panel.  The expression takes into account the sound and Alfv\'{e}n speed as well as the RMS velocity. We see that the filamentary structures are far more prone to gravitational fragmentation than the diffuse gas - these filamentary clouds are both magnetically and gravitationally supercritical. Interestingly, the critical line-mass appears to be rather constant along each spur and shows no significant variation with position in the disc with values of $(M/L)_\mathrm{crit}\sim6\times10^3-10^4\,\mathrm{M}_\odot\,\mathrm{pc}^{-1}$. \\
\indent In order to identify the scale of the instability, in Fig. \ref{fig5} we plot azimuthal energy spectra of the vertical B-field component in a cubic region with an edge length of 2~kpc for two initial magnetisations of the disc galaxy. In all simulations, the vertical magnetic field grows with time as indicated by the increase in the spectral energy. 
Particularly interesting is the wavenumber at which the energy peaks, because it gives the characteristic length scale for the buoyancy of the field. This wavenumber (normalised to the box size), $k$, can be estimated to be at $k\sim2-3$. Note that the spectrum for the weakly magnetised galaxy (initial plasma-$\beta=10$) peaks at a larger wavenumber as this galaxy evolves almost like the hydrodynamical model. These wavenumbers give characteristic scales of $L=0.7-1\,\mathrm{kpc}$ for the azimuthal perturbation, in good agreement with early theoretical studies \citep{Parker66} and previous numerical simulations \citep{Kim01,Kim02,Kim02b}.
\section{Discussion and Conclusion}
Our simulations show that gas accumulation along magnetic field lines does not lead to the formation of GMCs directly. Instead, kpc-scale coherent flows along buoyant magnetic field lines first lead to the formation of giant filaments by converging in magnetic valleys. The mass reservoir provided by this mechanism is sufficient to render the magnetic field magnetically supercritical. From an observational perspective these filaments resemble the spur-like features that have been noted in galaxies \citep{Ho09}. We note also that 
seven giant molecular filaments 
(GMFs) that have lengths on the order of 100~pc, total masses of $10^4 - 10^5 M_{\odot}$ 
exhibiting velocity coherence over their full length  are observed in the inter arm regions of our galaxy \citep[termed 'Giant Molecular Filaments',][]{Ragan14}.
 In short, the filaments in our simulated discs are magnetically supercritical, have low levels of 
turbulence, and have low critical mass per unit length. Whereas the initial conditions of previous studies were magnetically supercritical \citep{Kim01,Kim02,Mouschovias09}, we observe spur formation and fragmentation also for initially highly subcritical galaxies. \\
\indent We conclude that the Parker instability enables the formation of magnetically supercritical filaments 
out of an initially subcritical diffuse interstellar medium. H\textsc{i} flows into long magnetic valleys 
 and produces filamentary clouds that undergo subsequent gravitational fragmentation.  These fragments are fed by flows in the filaments, resulting in the 
 rapid accumulation of GMC mass clouds.
 Our results provide a  solution to the question of how 
to form clouds out of a highly magnetically subcritical ISM, posed half a century ago. We further emphasise that the behaviour uncovered by our simulations is not merely a transient event based on idealised initial conditions.  While we cannot follow subsequent star formation in these simulations, it will have the effect of  eventually dispersing the  GMCs back into a diffuse medium.   Galactic shear will rapidly comb out the fields and the cycle will begin again. 
\vspace{-0.5cm}
\section*{Acknowledgements} 
We thank the referee for a timely and insightful report. We further thank E. V\'{a}zquez-Semadeni,  B.  Elmegreen, and C. McKee for discussions. BK and RB thank for funding via the DFG priority program 1573 'The Physics of the Interstellar Medium' (BA 3706/3-2). Furthermore RB acknowledges additional funding from the DFG for this project via the grant BA 3706/4-1. RB and WS are supported by the DFG for the project via the grant BA 3706/15-1. The FLASH software was developed in part by the DOE NNSA ASC- and DOE Office of Science ASCR-supported Flash Center for Computational Science at the University of Chicago. REP is supported by a Discovery grant from NSERC - Canada. We thank the Gauss Centre for Supercomputing e.V. (www.gauss-centre.eu) for funding this project (ID: pr92pu) by providing computing time on the GCS Supercomputer SuperMUC at Leibniz Supercomputing Centre (www.lrz.de). 

\bibliography{astro} 

\bibliographystyle{mn2e}

\end{document}